\documentclass[twocolumn,amssymb, nobibnotes, aps, prxquantum, superscriptaddress,floatfix]{revtex4-2}

\usepackage{amsmath}
\usepackage{braket}
\usepackage{bm}
\usepackage{graphicx}
\usepackage{float} 
\usepackage{subfigure} 
\usepackage{color}
\usepackage[normalem]{ulem}
\usepackage[ruled,linesnumbered]{algorithm2e}
\usepackage{array}
\usepackage{comment}
\usepackage[utf8]{inputenc}
\usepackage[T1]{fontenc}
\usepackage[english]{babel}
\makeatletter
\addto\captionsenglish{\renewcommand{\selectlanguage}[1]{}}
\makeatother

\usepackage{hyperref}

\usepackage{tikz}
\usetikzlibrary{calc}

\definecolor{myblue}{RGB}{122,191,223}
\definecolor{myorange}{RGB}{228,164,123}

\begin{document}

\title{Revealing Physical Redundancy in the Two-Dimensional Fermi-Hubbard Model via Transferable Observable Reconstruction}

\author{Ao-Ning Wang}
\affiliation{Department of Physics and HK Institute of Quantum Science \& Technology,
The University of Hong Kong, Pokfulam Road, Hong Kong, China}
\affiliation{Hong Kong Branch for Quantum Science Center of Guangdong-Hong Kong-Macau Greater Bay Area, Shenzhen, China}
\author{Min-Quan He}
\email{minquan@connect.hku.hk}
\affiliation{Department of Physics, City University of Hong Kong,
Tat Chee Avenue, Kowloon, Hong Kong SAR, China}
\affiliation{City University of Hong Kong Shenzhen Research Institute, Shenzhen, Guangdong 518057, China}
\affiliation{Quantum Science Center of Guangdong-Hong Kong-Macao Greater Bay Area, Shenzhen, Guangdong 518045, China}
\author{Z. D. Wang}
\email{zwang@hku.hk}
\affiliation{Department of Physics and HK Institute of Quantum Science \& Technology,
The University of Hong Kong, Pokfulam Road, Hong Kong, China}
\affiliation{Hong Kong Branch for Quantum Science Center of Guangdong-Hong Kong-Macau Greater Bay Area, Shenzhen, China}

\begin{abstract}
The Fermi-Hubbard model provides a paradigmatic setting for studying strongly correlated quantum matter, where different observables are commonly used to probe charge, interaction, and spin correlations. In this work, we investigate whether these observables contain mutually transferable physical information beyond their apparent distinction. We quantify such physical redundancy through transferability tests among three representative observables of the two-dimensional Fermi-Hubbard model: total density~(N), double occupancy~(D), and spin-spin correlation~(S). Using a neural-network reconstruction framework, we find that the phase diagram of one observable can be reconstructed from another with accuracy close to self-reconstruction benchmarks, especially in trivial phase regimes. This transferability relies on correct physical labeling, persists across finite-temperature regimes, and remains robust under noisy inputs. Our results suggest that separate observables can carry a substantial fraction of one another's physical information, providing numerical evidence for observable-level redundancy in the two-dimensional Fermi-Hubbard system.
\end{abstract}

\maketitle

\section{Introduction}
The Fermi-Hubbard Model has long been regarded as a fundamental framework for studying Mott physics and understanding strongly correlated phenomena in lattice systems~\cite{FHM_importance_1, FHM_importance_2, FHM_importance_3}. It also experimentally serves as a cornerstone for the experimental realization of ultracold-atom quantum simulation in optical lattices~\cite{FHM_qs_1}, underscoring its central role in contemporary quantum research. Three most iconic and extensively studied physical observables of the 2D Fermi-Hubbard Model are total density, double occupancy, and spin-spin interaction. Each of these observables encodes distinct information about the system and plays a unique role in determining its macroscopic behaviors~\cite{property_importance_1, property_importance_2, property_importance_3, property_importance_4}. From a physical perspective, these observables are expected to be intrinsically correlated, and several widely accepted approximate relations have indeed been proposed over the past decades to connect them in specific limiting regimes. The general derivation of one observable from another, however, remains intrinsically intractable~\cite{theory1, relation_1, property_importance_4, relation_2}. In this work, we take a step beyond this theoretical limitation. We conduct a numerical transferability study, in which we employ the three observables to cross-reconstruct each other's phase diagram. This process not only empirically probes correlations between individual observables, but also provides a quantitative measure of the intrinsic physical information shared among them, thereby revealing the degree of physical redundancy present in the system. 

In this work, we adopt a numerical framework, an approach that has become increasingly prevalent in recent years with the rapid advancement of computational power and artificial intelligence~\cite{numerical_for_physics_1, numerical_for_physics_2, numerical_for_physics_3, numerical_for_physics_4}. Our central idea is to employ a carefully designed neural network to perform a series of regression tasks. In each task, one observable is provided as the input, while the neural network is expected to exploit all correlations possible and learn a mapping function to reconstruct another observable's phase diagram from the given input. After eliminating alternative sources of influence through shuffle and noise tests, the performance of this learned mapping reflects its reproducible reconstructability, and consequently, the amount of intrinsic physical information shared between the two distinct observables. Our results reveal a clear and consistent trend. Despite the fact that these observables occupy different levels in the information hierarchy in the sense of their distinct physical correlation lengths, the transferability tests achieve remarkably strong performance across all input-output pairs. The transferred mappings attain nearly the same level of accuracy as the corresponding self-mapping benchmarks, in which the target observable itself serves as the input for reconstructing its own phase diagram. This capability persists throughout the entire thermodynamic range considered, and deteriorates only in the vicinity of phase-critical regimes. These findings suggest that these seemingly independent observables are more than merely correlated. Instead, they exhibit substantial physical redundancy, implying that distinct observables may reside on a common higher-order many-body manifold that encodes shared physical information.

\begin{figure}
    \centering
    \includegraphics[width=1.0\linewidth]{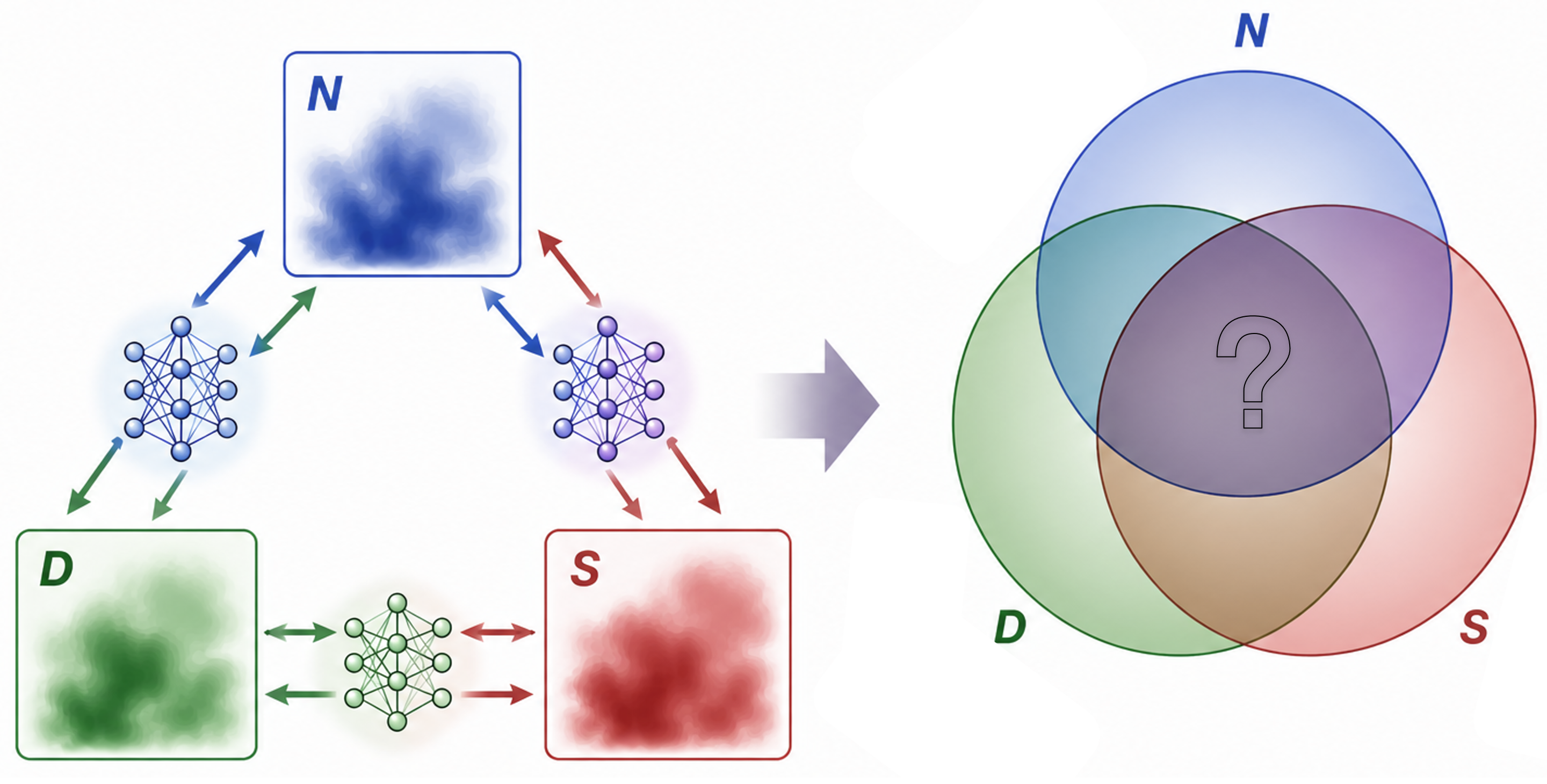}
    \caption{Physical redundancy in observables. Transferability mapping tests with a neural network provide an operational way to quantify shared physical information among distinct observables.}
    \label{figure_1}
\end{figure}

\section{The 2D Fermi-Hubbard Model}
The Fermi-Hubbard Model describes electron hopping behaviors within a lattice grid. The competing effects include neighboring hopping, local Coulomb interaction and chemical potential bias. There are therefore three competing terms within the system Hamiltonian, as shown in order in Eq \ref{fermion_Hamiltonian}~\cite{theory1, theory2, theory3, theory4}. $\sigma$ refers to spin direction and takes two values: $\uparrow$(up) and $\downarrow$(down). $r$ refers to a lattice site that can accommodate up to two opposite-spin electrons according to Pauli exclusion principle, and $\langle r,r' \rangle$ refers to a pair of spatially adjacent lattice sites. $c_{r\sigma}^\dagger/ c_{r\sigma}$ are the canonically anticommute fermion operators that create/annihilate a fermion with spin $\sigma$ at site $r$. $n$ values in the last two terms are the corresponding fermion numbers at certain lattice sites with certain spin directions. Note that $n$ can be easily written in fermion operators, but we choose to write out $n$ explicitly here because we want to show more clearly the physical meaning of each term. It can be easily seen that each of the three terms has a strength coefficient attached, weighting their corresponding competence in the Hamiltonian. $t$ is called the hopping strength, and it can be seen from the expression of the first term that it describes the hopping intensity of electrons across adjacent lattice sites. $U$ refers to local Coulomb repulsion energy, and the second term punishes a fully occupied lattice site with two opposite-spin electrons. Finally, chemical potential $\mu$ acts as a background bias that controls the total electron number.
The system in its general form is analytically unsolvable, meaning that it has no known closed-form solution and is exponentially hard to scale numerically with any existing methods. Therefore, in the main part of this study, we will simulate the system in a moderate setting: a single-band 2D 2$\times$2 lattice grid with four lattice sites, which is the basic building block of a 2D band system. It is simple but inspiring, with one plaquette capturing key 2D physical characteristics. At the end of Results section however, we provide a substantial sample feasibility test with several larger lattice architectures, which serves as a core clarification of finite-size effect. We will delay a detailed discussion until then.
\begin{equation}
    \label{fermion_Hamiltonian}
    \begin{aligned}
    \hat{H} = &-t \sum_{\sigma \in \{\uparrow,\downarrow\}} \sum_{\langle r,r' \rangle} \bigl( c_{r\sigma}^\dagger c_{r'\sigma} + c_{r'\sigma}^\dagger c_{r\sigma} \bigr) \\ &+ U \sum_r n_{r\uparrow} n_{r\downarrow} \\ &- \mu \sum_{r}\sum_{\sigma} n_{r\sigma}
    \end{aligned}
\end{equation}

Before the transferability test, we need to build our central dataset with a simulator. To do that, we do not prefer the Hamiltonian in its fermion-operator form. Instead, we want an equivalent spin model, which will be directly deployable on our device. Therefore, our next step would be to make adjustments to the fermion-operator Hamiltonian with Jordan-Wigner transformation. This means an effective translation from fermion operators into Pauli spin operators, following the formula in Eq \ref{JW_transformation}. $m$ here refers to any unique electron state with a unique pair of location and spin. It is quite noticeable that different ordering of modes can lead to different implementation practices in simulation while they effectively share the same physics due to basis change symmetry in Jordan-Wigner transformation~\cite{order}. A detailed demonstration of our ordering choice can be found in Methods.
\begin{equation}
    \label{JW_transformation}
    \begin{aligned}
        c_m&=\left(\prod_{k=0}^{m-1} Z_k\right)\frac{1}{2}\left(X_m+iY_m\right)\\
        c_m^\dagger&=\left(\prod_{k=0}^{m-1} Z_k\right)\frac{1}{2}\left(X_m-iY_m\right)
    \end{aligned}
\end{equation}
After all the conversion between operators, we finally arrive at a convenient spin-form Hamiltonian in Pauli spin operators in the following form~\cite{theory1, theory2, theory3, theory4}:
\begin{equation}
\label{spin_Hamiltonian}
\begin{aligned}
    \hat{H} = &-\frac{t}{2} \sum_{\langle r,r' \rangle} \sum_{\sigma}(X_{i}X_{j} + Y_{i}Y_{j})\Pi_{k=i+1}^{j-1}Z_{k}\\ &+ \frac{U}{4} \sum_{r} (1 - Z_{u(r)} - Z_{d(r)} + Z_{u(r)}Z_{d(r)})\\ &- \frac{\mu}{2}\sum_{r}(2 - Z_{u(r)} - Z_{d(r)})
\end{aligned}
\end{equation}
where $\sigma$ now corresponds to either of the two spin directions u(up) and d(down), $\langle r,r' \rangle$ still represents a pair of adjacent lattice sites, and $(i,j)$ refers to either of the two pairs of modes with a same spin direction within a $\langle r,r' \rangle$ pair. Note that this Hamiltonian now includes long-range interactions and series multiplication.

The three physical observables we are interested in are total density, double occupancy and spin-spin interaction at $(\pi, \pi)$ respectively. Total density reflects on the total number of electrons induced, double occupancy reflects on the number of sites that are fully occupied with two opposite-spin electrons, and spin-spin interaction reflects on neighboring spin alignment. In terms of correlation length and therefore information hierarchy, total density is only related to a local measurement, while double occupancy includes local interaction within a single lattice site. Spin-spin interaction on the other hand is subject to long-range interaction, and in its least form includes neighboring lattice sites.
Together they cover most of the significant charge, interaction, and spin-correlation characteristics of the system. Their corresponding spin operators after Jordan-Wigner transformation are shown in Eq \ref{total_density_operator}, \ref{double_occupancy_operator} and \ref{spin_spin_interaction_operator}~\cite{theory1, theory2, theory3, theory4}, where $(r-r')$ refers to $(x_{r}-x_{r'})+(y_{r}-y_{r'})$ is a count of coordinate distance in the unit of lattice parameter. The spin-spin interaction observable used in this work is a staggered longitudinal spin-spin correlation. Instead of evaluating the full vector product $\mathbf{S}_{\mathbf r}\cdot \mathbf{S}_{\mathbf r'}$, we retain only the longitudinal component $S^z_{\mathbf r}S^z_{\mathbf r'}$ and include a staggered phase factor between lattice sites. The final operator is then obtained by summing this weighted longitudinal correlation over lattice-site pairs. We present the observation results in an average manner, where expectation values from all three properties would be averaged to each lattice site. In a 2$\times$2 2D square lattice grid for example, there will be 4 lattice sites.
\begin{align}
    \hat{n} &= \frac{1}{N_{sites}} \sum_{m=1}^{N_{modes}}\frac{1 - Z_{m}}{2} \label{total_density_operator}\\
    \hat{d} &= \frac{1}{N_{sites}} \sum_{r}\frac{(1 - Z_{u(r)})(1 - Z_{d(r)})}{4} \label{double_occupancy_operator}\\
    \hat{s} &= \frac{1}{N_{sites}} \sum_{r}\sum_{r'}\frac{1}{16}(-1)^{(r-r')}(Z_{d(r)} - Z_{u(r)})(Z_{d(r')} - Z_{u(r')})\label{spin_spin_interaction_operator}
\end{align}

\section{The Dataset and Neural Network}
\begin{figure}
    \centering
    \subfigure[]{ \includegraphics[width=0.9\linewidth]{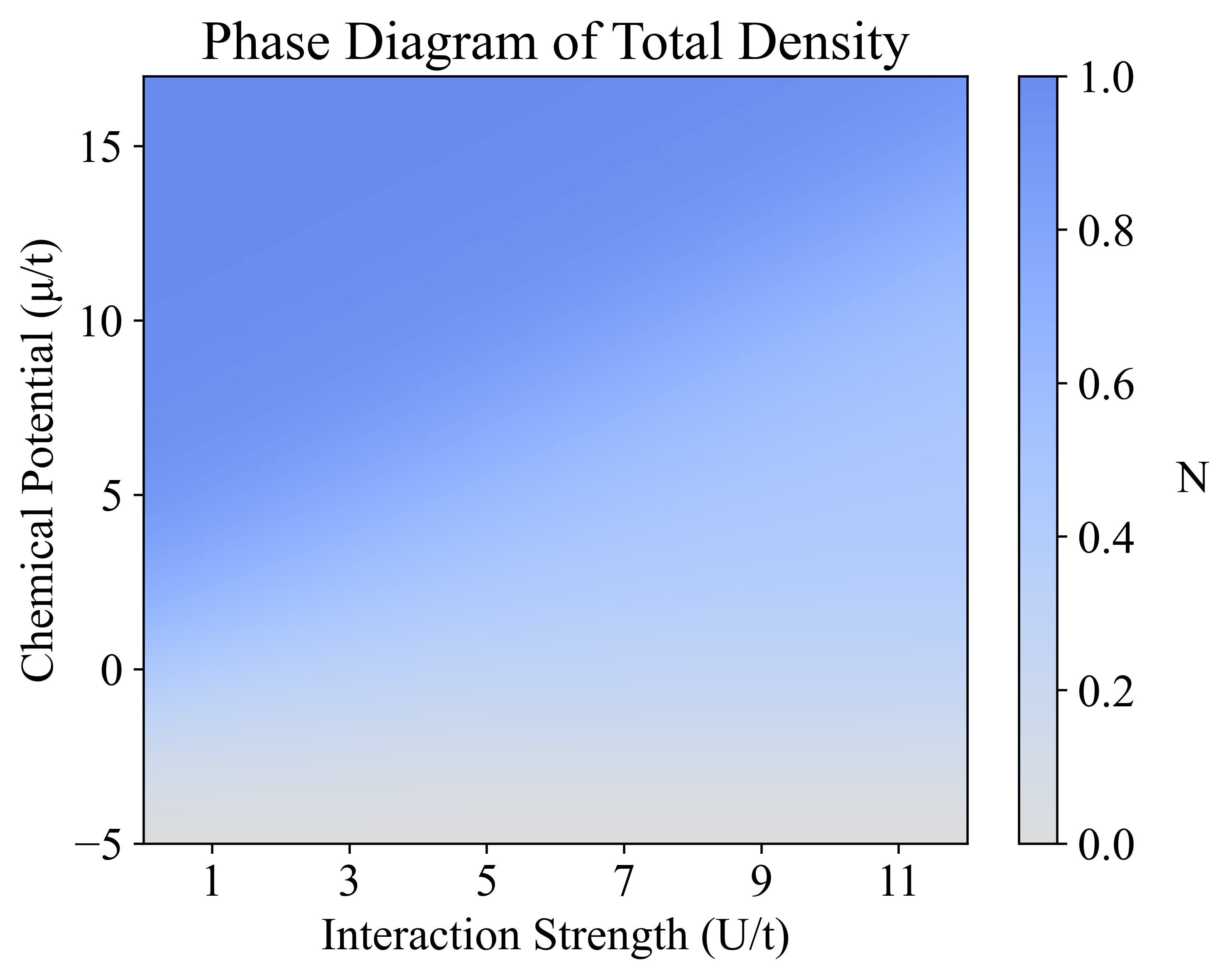}}
    \subfigure[]{ \includegraphics[width=0.9\linewidth]{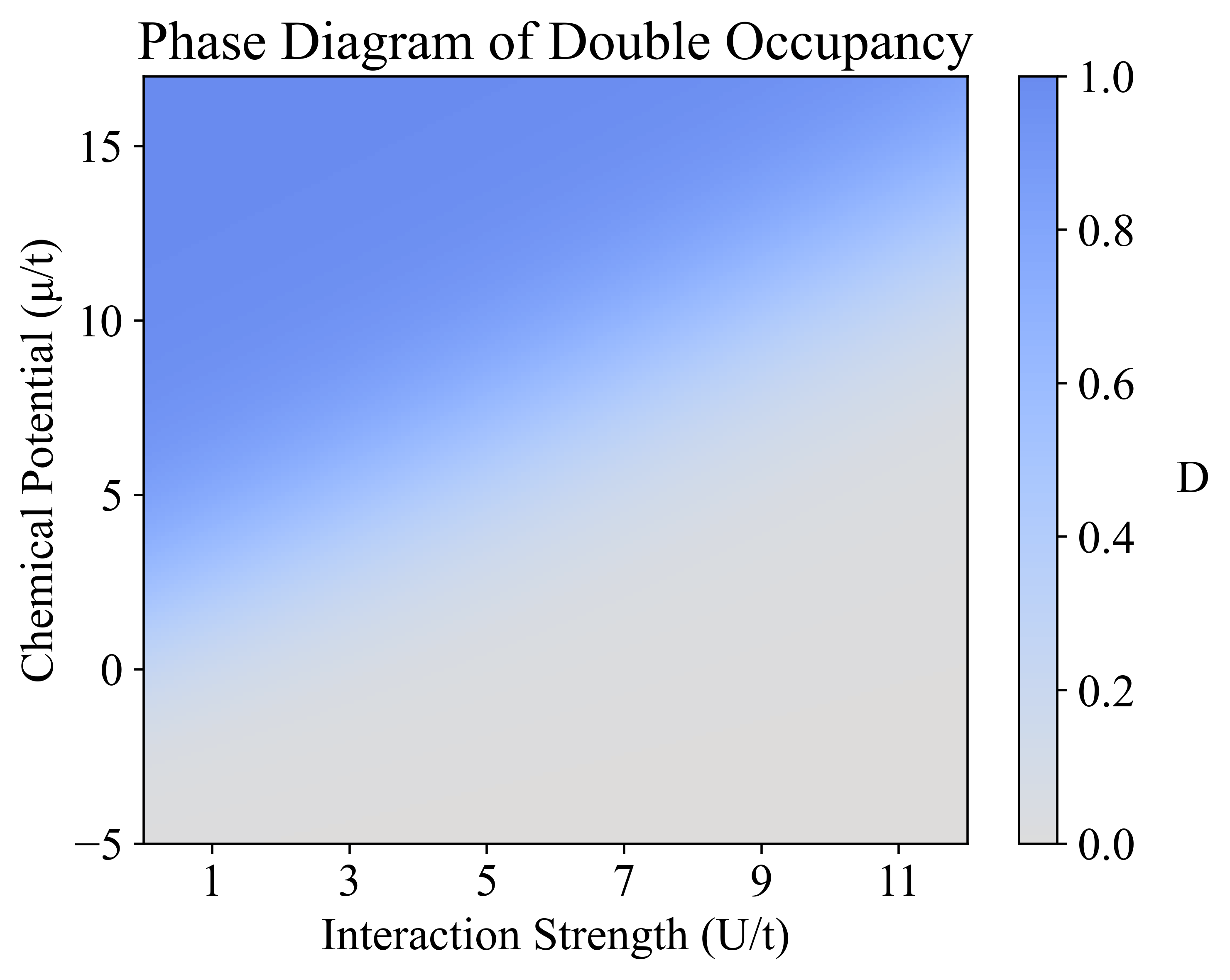}}
    \subfigure[]{ \includegraphics[width=0.9\linewidth]{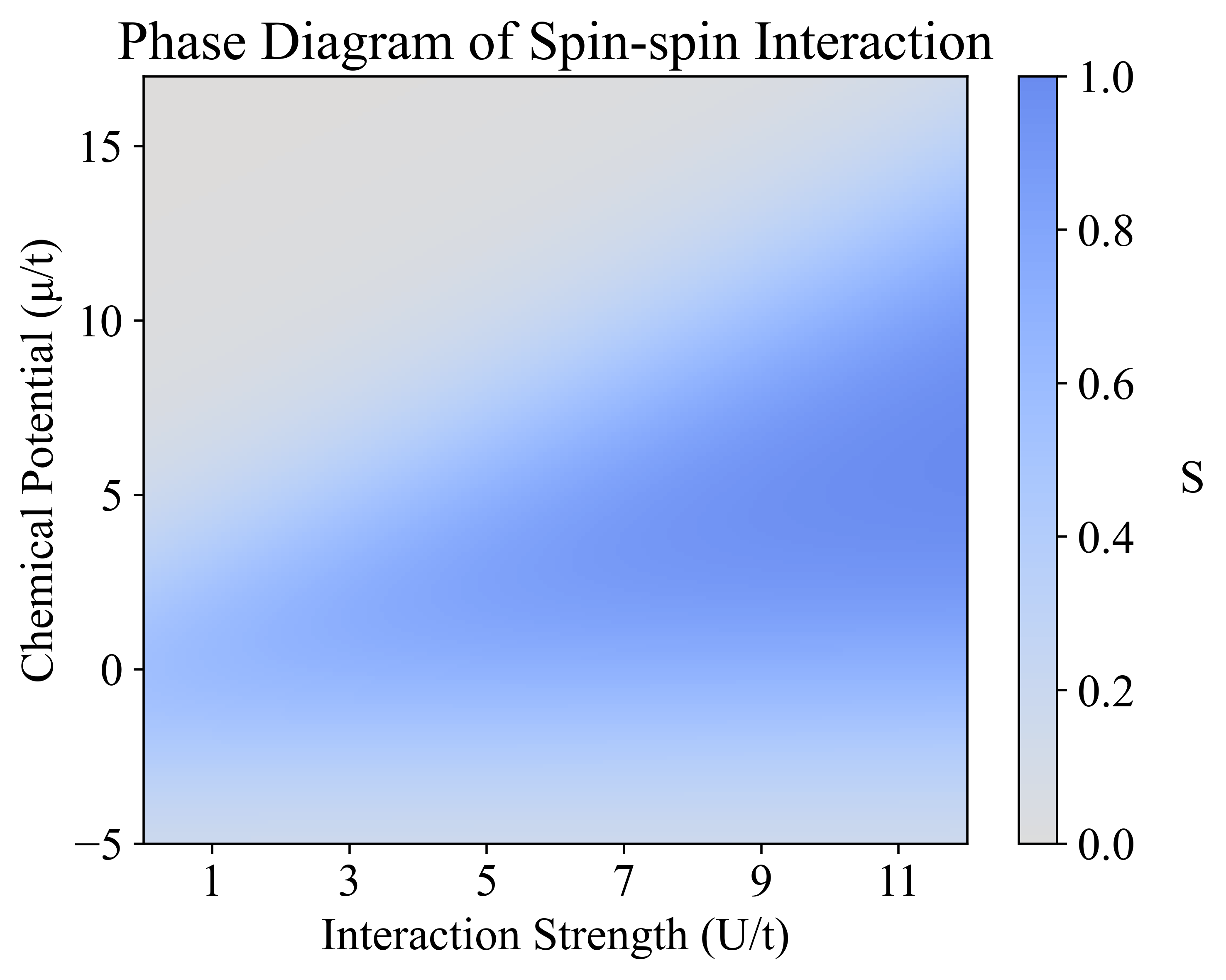}}
    \caption{Sample phase diagrams of (a) total density, (b) double occupancy, and (c) spin-spin interaction. Each data point represents an expectation evaluation taken with respect to the Gibbs State at the sample temperature. All three diagrams are min-max normalized for display.}
    \label{phase_diagrams}
\end{figure}

As we have mentioned, our central dataset, which will be used in the main part of this work, comes from quantum simulation with the above Hamiltonian and observables. For each of the three observables, we evaluate its expected values according to three variables $(T,U,\mu)$. In particular, $T$ is used as a scanning parameter. That is to say, each piece of data~(in the form of a phase diagram) in our dataset corresponds to a unique thermodynamic temperature $T$, meaning that all data points included in this piece of data are evaluation results upon the system Gibbs State at this specific temperature. Gibbs State at temperature $T$ is defined with $k_{\mathrm{B}}=1$ as
\begin{equation}
    \rho(T) = \frac{e^{-\hat{H}/T}}{\mathrm{Tr}\left(e^{-\hat{H}/T}\right)}
    \label{Gibbs}
\end{equation}
The other two variables $(U,\mu)$ on the other hand act as the axes in each phase diagram. Details of this treatment can be seen in Methods. Following this process, we complete our dataset for all three observables, and we display a piece of sample data for each of them in Figure \ref{phase_diagrams}. Critical crossover regions can be seen on all of the phase diagrams, which originate from different physical mechanisms and serve as different physical indicators. This is not the main focus of this work but we will briefly discuss it in Results.

For the neural network, we choose a multi-layer perceptron~(MLP) network~\cite{MLP}. As we have mentioned, the preliminary aim of this transferability test is to numerically capture intrinsic physical redundancy by training, which could very possibly include highly complex non-linear functions. Moreover, we find in practice that when we try to reconstruct observable B's phase diagram from observable A, we better supervise this process with observable A's phase diagram as well. Otherwise, the neural network tends to get lost without any reference. Note that this supervision process includes no information leakage and is fully legitimate. Details can be seen in Methods. It can be seen from the above analysis that the choice of our neural network must at least satisfy these two conditions: 1) the work space must be broad enough to simultaneously include data from separate observables; and 2) the expressibility must be strong enough to allow sufficient capture of intrinsic physical correlations. Therefore we consider an MLP network with deep hidden layers appropriate in our case.

To prepare for training, we apply a trainable encoder $E_{\varphi}$ in EDSR style~\cite{EDSR} to our input observable A, which would be presented in a clipped-phase-diagram form, also with axes, details can be seen in Methods. Data points in the input would be encoded into vectors, which would later be used non-linearly in mapping. Borrowing ideas from Convolutional Neural Network studies~\cite{CNN}, we further do a local concatenation of vectors before mapping, as shown in Eq \ref{concatenation}, so that we cover smooth spatial trends across our input. $\mathbf{z}_{m,n}$ is the final concatenated vector at input coordinate $(m,n)$, and it is a concatenation result of initial self and the eight nearest vectors surrounding it on a $3\times3$ square grid. Note that this grid is on phase diagram rather than lattice.
\begin{equation}
    \label{concatenation}
    \mathbf{z}_{m,n} = Concat(\{\mathbf{z}_{m+i,n+j}\}_{i,j=-1,0,1})
\end{equation}

\begin{figure}
    \centering
    \includegraphics[width=0.8\linewidth]{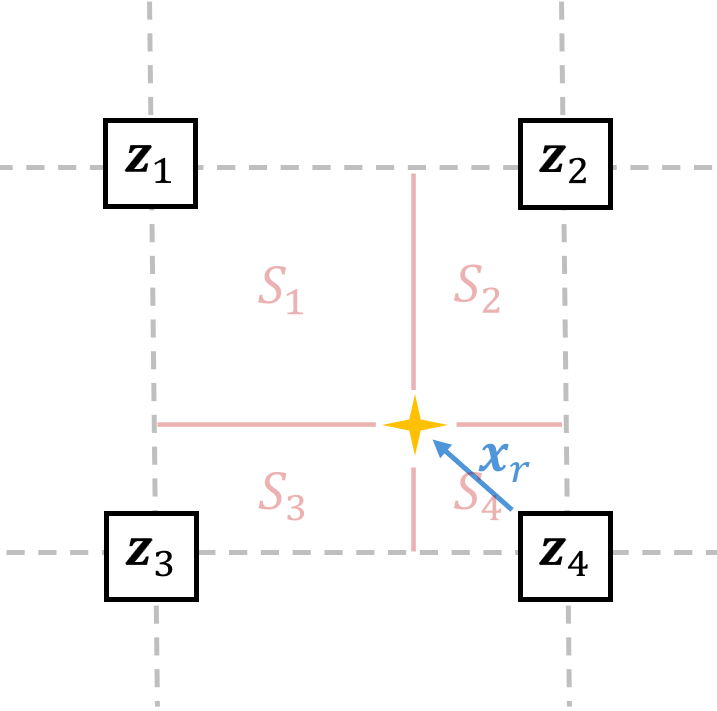}
    \caption{Graph demonstration of decoding function $f_{\theta}$. $\mathbf{z}_{4}$ is in this case $\mathbf{z}_{near}$ if only one nearest vector is considered. $S_{1}$, $S_{2}$, $S_{3}$ and $S_{4}$ are the corresponding relative areas between target coordinate and the nearest four corners on an input $2\times2$ square grid.}
    \label{demo}
\end{figure}

With vectors at all input coordinates well prepared, we apply a trainable decoding function $f_{\theta}$ to reconstruct our output observable B's phase diagram. To do that, we need to map our input to every coordinate on B's phase diagram respectively. If target coordinate is $\mathbf{x}$, this is done by taking into $f_{\theta}$ as arguments input vector $\mathbf{z}_{near}$ located next to $\mathbf{x}$~(because input diagram is clipped and there may not be a $\mathbf{z}_{exact}$ exactly located on $\mathbf{x}$) and the relative coordinate distance $\mathbf{x}_{r}$ between them. This treatment is similar to that has been done in~\cite{LIIF}. Demonstrations can be seen in Eq \ref{decoding_function} and Figure \ref{demo}. However, this implementation may cause a sudden jump of mapping result when we spatially sweep across target coordinates, because the judgment of nearest vector can suddenly change at the midpoint of two vectors. To solve this issue, instead of considering only the top one nearest vector, we take into account all the top four nearest vectors in every mapping. That is basically referring to the four corners on a $2\times2$ square grid which stores the target coordinate. We still apply $f_{\theta}$ respectively onto each of the four vectors, only this time, we take a weighted average of the four mapping results to be our final mapped $\mathbf{x}$, as illustrated in Eq \ref{weighted_average} and Figure \ref{demo}. The weightings trace back to comparison of relative areas between target coordinate and each of the four corners. The core trainable parameter $\theta$ in the decoding function is just in the form of our chosen fully connected MLP.
\begin{equation}
    \label{decoding_function}
    Mapping Result = f_{\theta}(\mathbf{z}_{near},\mathbf{x}_{r})
\end{equation}
\begin{equation}
    \label{weighted_average}
    Mapped(\mathbf{x}) =  \Sigma_{t \in \{1,2,3,4\}} \, \frac{S_{t}}{S_{1}+S_{2}+S_{3}+S_{4}} \, f_{\theta}(\mathbf{z}_{t},\mathbf{x}_{r_{t}})
\end{equation}

During training, we take similar approach to~\cite{LIIF} and consistently compare mapping results with observable B's true phase diagrams within our dataset. We perform $L_{1}$ loss evaluation and Adam optimization~\cite{Adam} during each epoch to update $\theta$ and $\varphi$ simultaneously so that both encoder $E_{\varphi}$ and decoding function $f_{\theta}$ can eventually be optimized. This whole architecture including local convolution and coordinate inferring will guarantee that spatial correlations on phase diagrams are best identified and encoded, so that at the end of training, we get the best transferred mapping results possible.

\section{Results}
We will display our results following a two-step logic. We will first verify our primary argument, that there indeed is redundancy among our three target observables, before we carry on to explore possible scenarios where observables decouple and this redundancy collapses. As we have discussed above, the three observables can be categorized into different information hierarchy in the sense of their correlation lengths. This would create two different directions of information flow in transferability tests, namely from low to high
\begin{equation}
    N \rightarrow D \rightarrow S
\end{equation}
and from high to low
\begin{equation}
    S \rightarrow D \rightarrow N
\end{equation}
We will try to cover both directions of information flow whenever possible to ensure comprehensiveness.

\subsection{Transferability Matrix}
The first table we would like to present are results from our comprehensive transferability test, where we have recorded mapping performance in all possible "from A map to B" experiments. Values in Table \ref{psnr_matrix} (a) stand for an average Peak Signal Noise Ratio~(PSNR) performance in that particular "A to B" experiment, details can be seen in Methods. Values in Table \ref{psnr_matrix} (b) on the other hand stand for a comparison of Mean Squared Error~(MSE) between each "A to B" and the corresponding "B to B" experiment. PSNR and MSE used here are defined as
\begin{align}
     MSE &= \frac{1}{m^{2}} \, \Sigma^{m-1}_{i = 0} \Sigma^{m-1}_{j = 0} \, [p(i,j) - t(i, j)]^{2} \\
    PSNR &= -10 \, log_{10}(MSE)
\end{align}
where each phase diagram we use as ground truth consists of $m\times m$ data points in total~($m$ different $U$ and $m$ different $\mu$ values), and $p(i,j)$ and $t(i,j)$ are the mapped and true values at data point $(i,j)$ respectively.

This transferability matrix shows a consistent cross-observable reconstruction performance. All "A to B" transferred experiments have achieved comparable PSNR level to the corresponding "B to B" self mapping benchmark, with a maximum MSE increase ratio barely constrained below 2. This means that in the case where our task is to reconstruct B's phase diagram, observable A has nearly the same physical information level compared to observable B itself. It is also important to point out that this transferability does not apparently depend on information hierarchy. If transferability happened to favor a particular information flow direction only, then we should argue instead that information hierarchy dominates in this process. Our results here on the contrary suggest that these seemingly distinct observables share a similar higher order manifold where information transfer happens in both directions.

\begin{table*}[t]
\centering

\caption{
Transferability analysis among the three observables:
total density ($N$), double occupancy ($D$),
and spin-spin interaction ($S$). (a) The mapping performance PSNR matrix, values are rounded to 2 decimal places in dB. (b) mapping MSE matrix, each value displayed is the MSE value yielded from an "A to B mapping" divided by the MSE value yielded from a "B to B mapping". That is, the MSE ratio between transferred and self mapping.
}

\vspace{0.5em}


\begin{minipage}[t]{0.49\textwidth}
\centering

\textbf{(a)} 

\vspace{0.6em}

\begin{tikzpicture}[scale=0.55]

\def\W{4}
\def\H{2.5}

\fill[myblue,opacity=0.3]
(0,0)
rectangle
(4*\W,4*\H);

\foreach \x in {0,1,2,3,4}
{
    \draw[white,line width=1pt]
    (\x*\W,0)
    --
    (\x*\W,4*\H);
}

\foreach \y in {0,1,2,3,4}
{
    \draw[white,line width=1pt]
    (0,\y*\H)
    --
    (4*\W,\y*\H);
}

\draw[line width=1pt,black]
(0.2,3.9*\H)
--
(3.8\W,3.1*\H);

\node[font=\bfseries\large]
at (0.35*\W,3.3*\H)
{From};

\node[font=\bfseries\large]
at (0.75*\W,3.7*\H)
{To};

\node[font=\bfseries\Large] at (1.5*\W,3.5*\H) {N};
\node[font=\bfseries\Large] at (2.5*\W,3.5*\H) {D};
\node[font=\bfseries\Large] at (3.5*\W,3.5*\H) {S};

\node[font=\bfseries\Large] at (0.5*\W,2.5*\H) {N};
\node[font=\bfseries\Large] at (0.5*\W,1.5*\H) {D};
\node[font=\bfseries\Large] at (0.5*\W,0.5*\H) {S};

\node[font=\normalsize] at (1.5*\W,2.5*\H) {40.45};
\node[font=\normalsize] at (2.5*\W,2.5*\H) {40.55};
\node[font=\normalsize] at (3.5*\W,2.5*\H) {41.79};

\node[font=\normalsize] at (1.5*\W,1.5*\H) {39.15};
\node[font=\normalsize] at (2.5*\W,1.5*\H) {43.53};
\node[font=\normalsize] at (3.5*\W,1.5*\H) {43.22};

\node[font=\normalsize] at (1.5*\W,0.5*\H) {39.91};
\node[font=\normalsize] at (2.5*\W,0.5*\H) {40.73};
\node[font=\normalsize] at (3.5*\W,0.5*\H) {44.15};

\end{tikzpicture}

\end{minipage}
\hfill
\begin{minipage}[t]{0.49\textwidth}
\centering

\textbf{(b)} 

\vspace{0.6em}

\begin{tikzpicture}[scale=0.55]

\def\W{4}
\def\H{2.5}

\fill[myorange,opacity=0.3]
(0,0)
rectangle
(4*\W,4*\H);

\foreach \x in {0,1,2,3,4}
{
    \draw[white,line width=1pt]
    (\x*\W,0)
    --
    (\x*\W,4*\H);
}

\foreach \y in {0,1,2,3,4}
{
    \draw[white,line width=1pt]
    (0,\y*\H)
    --
    (4*\W,\y*\H);
}

\draw[line width=1pt,black]
(0.2,3.9*\H)
--
(3.8\W,3.1*\H);

\node[font=\bfseries\large]
at (0.35*\W,3.3*\H)
{From};

\node[font=\bfseries\large]
at (0.75*\W,3.7*\H)
{To};

\node[font=\bfseries\Large] at (1.5*\W,3.5*\H) {N};
\node[font=\bfseries\Large] at (2.5*\W,3.5*\H) {D};
\node[font=\bfseries\Large] at (3.5*\W,3.5*\H) {S};

\node[font=\bfseries\Large] at (0.5*\W,2.5*\H) {N};
\node[font=\bfseries\Large] at (0.5*\W,1.5*\H) {D};
\node[font=\bfseries\Large] at (0.5*\W,0.5*\H) {S};

\node[font=\normalsize] at (1.5*\W,2.5*\H) {1.00};
\node[font=\normalsize] at (2.5*\W,2.5*\H) {1.99};
\node[font=\normalsize] at (3.5*\W,2.5*\H) {1.72};

\node[font=\normalsize] at (1.5*\W,1.5*\H) {1.35};
\node[font=\normalsize] at (2.5*\W,1.5*\H) {1.00};
\node[font=\normalsize] at (3.5*\W,1.5*\H) {1.24};

\node[font=\normalsize] at (1.5*\W,0.5*\H) {1.13};
\node[font=\normalsize] at (2.5*\W,0.5*\H) {1.91};
\node[font=\normalsize] at (3.5*\W,0.5*\H) {1.00};

\end{tikzpicture}

\end{minipage}

\label{psnr_matrix}

\end{table*}

\subsection{Shuffle and Noise Test}
To further exclude other influencing factors, we present our results from shuffle and noise tests here. With these two tests, we try to exclude these two possibilities respectively: 1. transferability comes from MLP trying to exploit numerical fluctuations rather than physical redundancy; and 2. transferability comes from an overfitting in mapping from an observable to another.

\begin{figure}
    \centering
    \subfigure[]{ \includegraphics[width=1.0\linewidth]{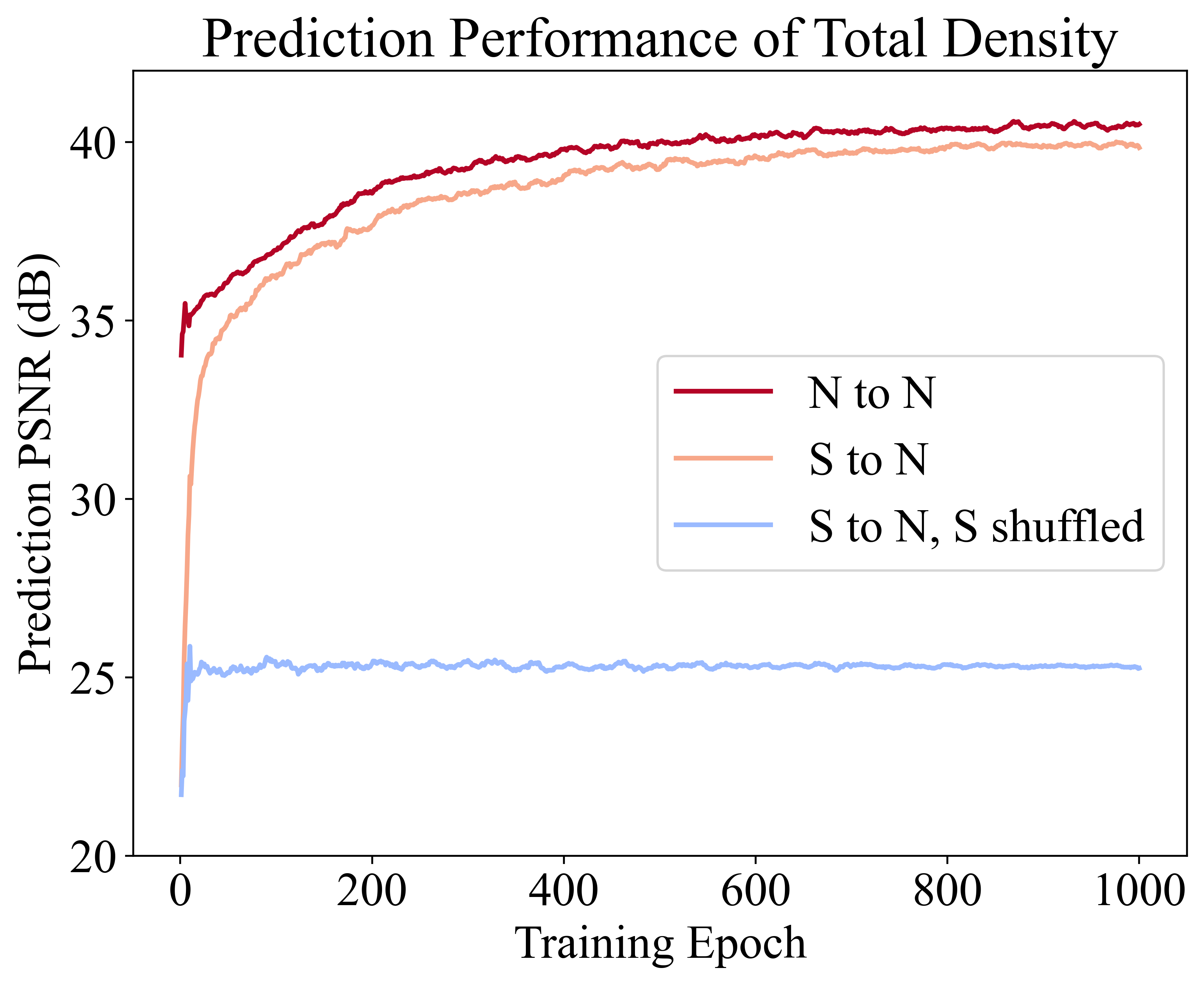}}
    \subfigure[]{ \includegraphics[width=1.0\linewidth]{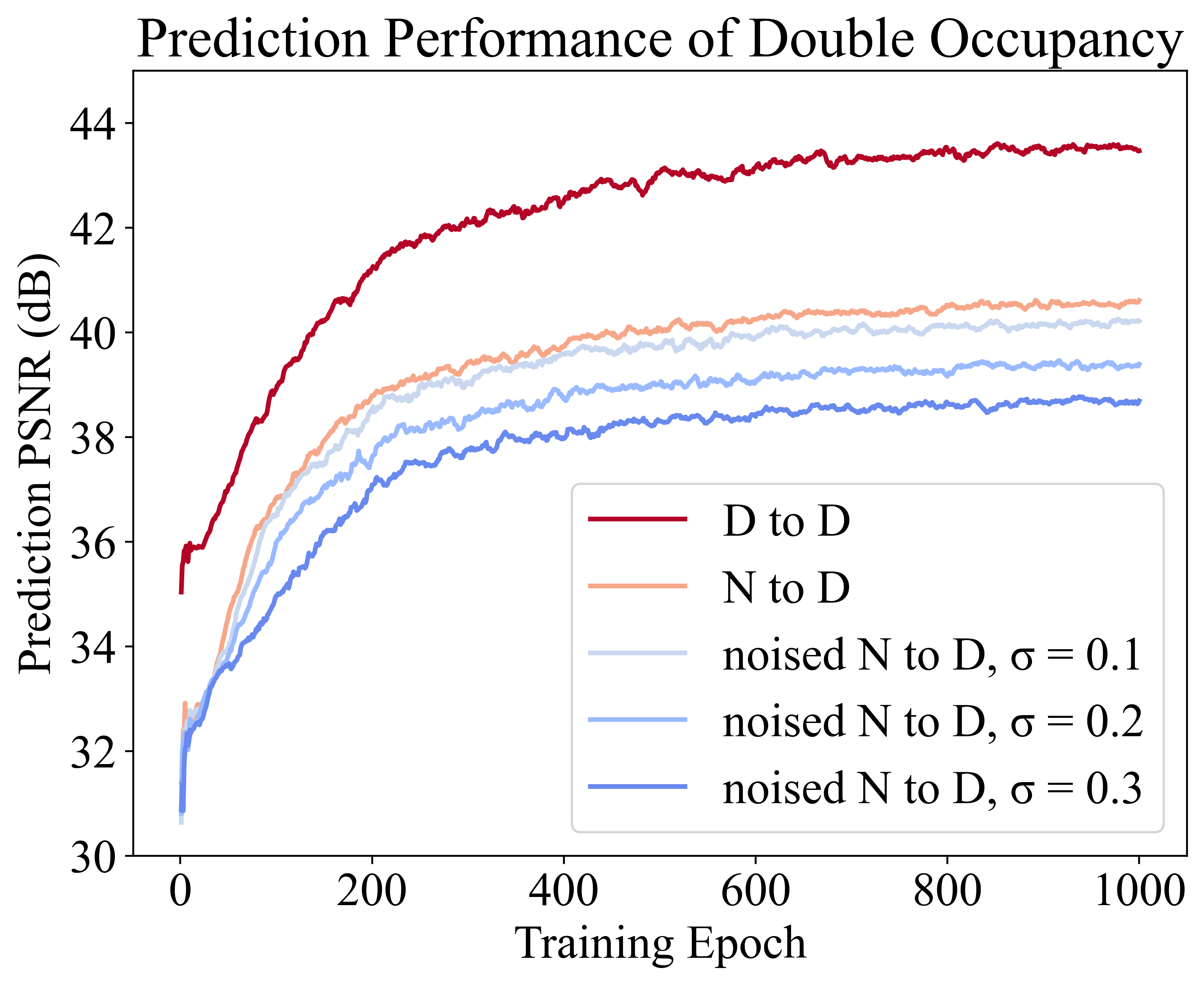}}
    \caption{Shuffle and noise tests. (a) Three separate mapping attempts of total density from itself, correct spin-spin interaction and shuffled spin-spin interaction respectively. (b) Five separate mapping attempts of double occupancy from itself, precise total density and three different-level noised total densities respectively.}
    \label{shuffle_noise_test}
\end{figure}

To exclude the first possibility, we need to create bait fluctuations which MLP can exploit but actually have no physical meanings. To do so, we add an extra shuffle operation prior to an ordinary mapping experiment "from S to N". Originally in our dataset, each piece of sample data contains three phase diagrams, corresponding to three observables evaluated upon the same system Gibbs State at the same temperature $T_{i}$. Consider only the two observables we are interested in this test, each piece of sample data is in the following form
\begin{equation}
(N_{T_{i}} , S_{T_{i}})
\end{equation}
Our ordinary mapping job is then to map to this $N_{T_{i}}$ from this $S_{T_{i}}$. What we try to do now is, we try to shuffle spin-spin interaction phase diagrams randomly across the whole dataset, so that the label between it and total density is destroyed. A piece of data now becomes
\begin{equation}
(N_{T_{i}} , S_{T_{j}})
\end{equation}
While preserving spatial relationship within every $S_{T_{j}}$ phase diagram to make sure it is still physically reasonable, the collapse of correct labeling should drastically increase the difficulty to map from it to $N_{T_{i}}$ because these two phase diagrams are now evaluation results from different temperatures and are supposed to have limited physical connections. Results in Figure \ref{shuffle_noise_test} (a) support this expectation, where it can be seen that "S to N" is intuitively more difficult than "N to N" at the beginning of training, but given correct physical labeling and therefore meaningful redundancy, the transferred mapping rapidly catches up as epoch increases. However, once S is shuffled and correct labeling is destroyed, the transferability attempt from S to N immediately fails. This behavior indicates that it is physical redundancy that MLP exploits, rather than arbitrary numerical fluctuations.

To exclude the second possibility, we need to create some artificial noise to test the robustness of mapping. Theoretically, a favorable mapping out of redundancy should depend on a local neighboring region, where the correspondence relationship is exploited out of a smooth trend. In other words, a Fourier filtering test on this mapping should favor low-pass characteristics, and small random noise in the input observable should not catastrophically break down this mapping. To verify that, we add Gaussian noise $G$ to our input prior to an ordinary "from N to D" mapping experiment. For each noise level $\sigma$ we apply, a self-proportional Gaussian noise is added to every data point $n$ on every $N$ phase diagram,
\begin{equation}
n_{noised} = clip\{n[1+G(0,\sigma^{2})],0,2\}
\end{equation}
The clip between $0$ and $2$ ensures that the noised total density phase diagrams stay strictly physical. Note that this process is done to the real values of $n$ rather than the sigmoid ones, details can be seen in Methods. We test four different noise levels
\begin{equation}
\sigma=0,0.1,0.2,0.3
\end{equation}
and the transferability results are shown in Figure \ref{shuffle_noise_test} (b). As expected, as noise level increases, mapping performance decreases monotonically due to information loss. However, even up to $\sigma=0.3$, this transferred mapping still exhibits a reasonable learnability, with a well constrained decrease in performance. This indicates that information needed for transferred reconstruction does not heavily depend on fine details of input, which helps to reduce the concern that transferred mappings originate from meaningless overfitting.

These two tests together justify that the transferability we see in Table \ref{psnr_matrix} has a physical origin rather than a purely numerical one. They suggest that it is intrinsic physical redundancy that enables accurate mappings between separate observables in both information-flow directions. We next examine the regimes in which this redundancy weakens and the observables decouple.

\subsection{Finite-Temperature Dependence}
The first natural question is whether the observed redundancy depends on thermodynamic temperature. As we mentioned above, our dataset comes from expectation evaluations upon the system Gibbs State at different temperatures, which means we are not evaluating the observables upon a pure state, instead, we always evaluate a thermodynamically mixed state. When temperature is sufficiently low, the Gibbs State would be very close to the system ground state, while when temperature is high, the Gibbs State would be highly mixed with different excited states. It is therefore expected that mapping performance, for both self-mapping and transferred mapping, should vary with temperature.

\begin{figure}
    \centering
    \includegraphics[width=1.0\linewidth]{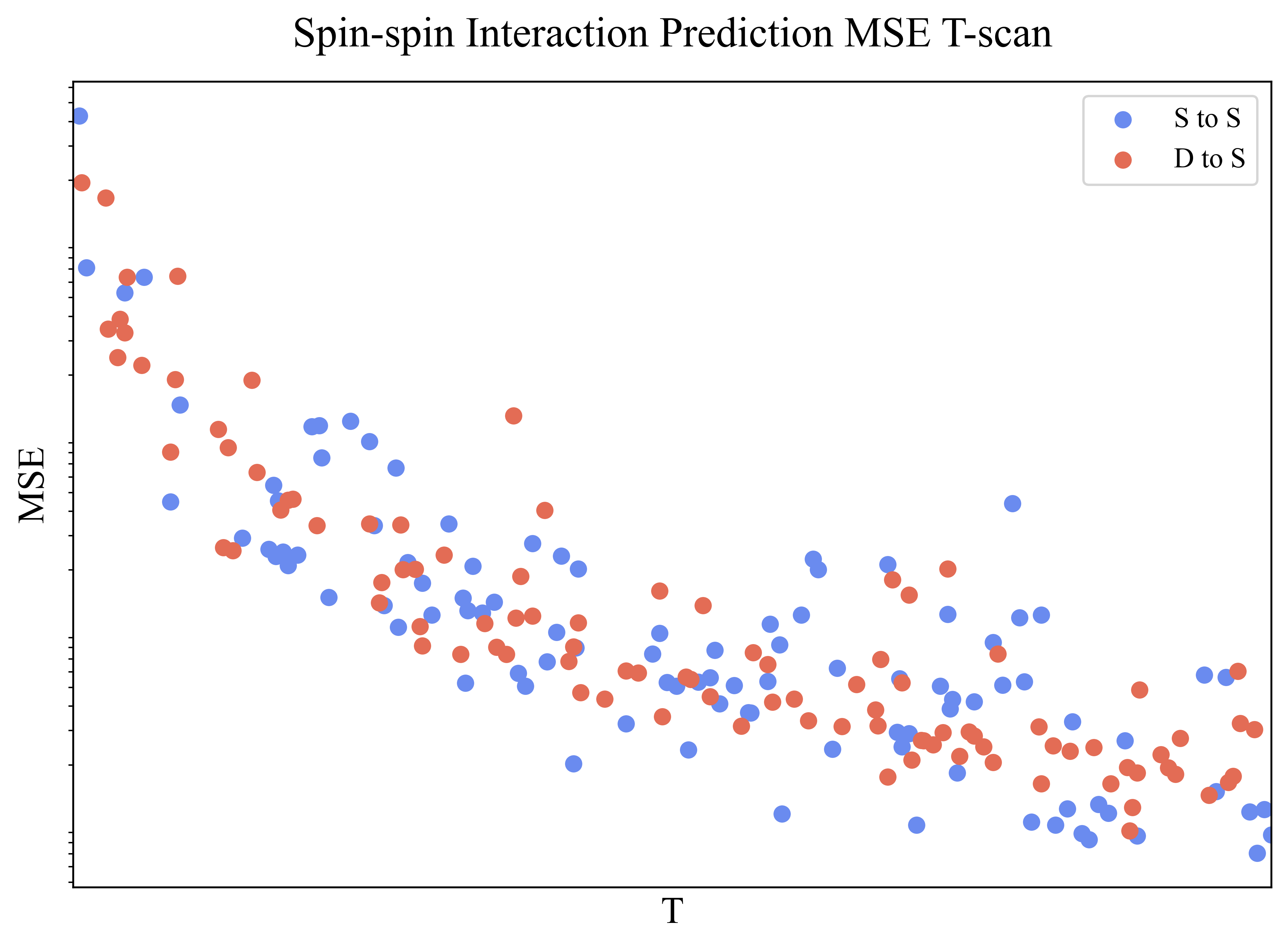}
    \caption{Mapping performance to spin-spin interaction scanned within the whole temperature range. Y axis is mapping MSE displayed in logarithmic scale. Results from both S to S and D to S attempts are displayed, similar trend can be seen in both groups that thermodynamically states closer to ground state are more difficult to capture.}
    \label{T_scan}
\end{figure}

Following this logic, we separate mapping error analysis according to ascending system temperatures in two reconstruction tests targeting spin-spin interaction, and display results in Figure \ref{T_scan}. It can be seen that overall mapping performance indeed depends strongly on temperature, and reconstruction becomes more difficult when the Gibbs state approaches the ground-state regime. This originates largely from the fact that transitions are much sharper around the ground state, which makes phase distinguishing harder for the MLP. However, the important thing to notice here is, we do not see clear performance gap between self mapping~("S to S") and transferred mapping~("D to S") throughout the temperature range studied. It seems that although it is well acknowledged that temperature plays a crucial role in governing magnetic phase transitions in the Fermi-Hubbard system~\cite{temperature_magnetic_1,temperature_magnetic_2}, the redundancy considered here does not vanish abruptly when thermodynamic temperature scans across critical points. 

\subsection{Phase Analysis}
Another important concern is the phase distribution of redundancy. As we have shown above, there are different phase regions in all three observables' phase diagrams. It is within expectation that physical information distribution would be different in separate regions, and therefore redundancy should also be dependent. Here we take sample data from our test set to do an error phase analysis.

\begin{figure*}
    \centering   
    \subfigure[]{ \includegraphics[width=0.34\linewidth]{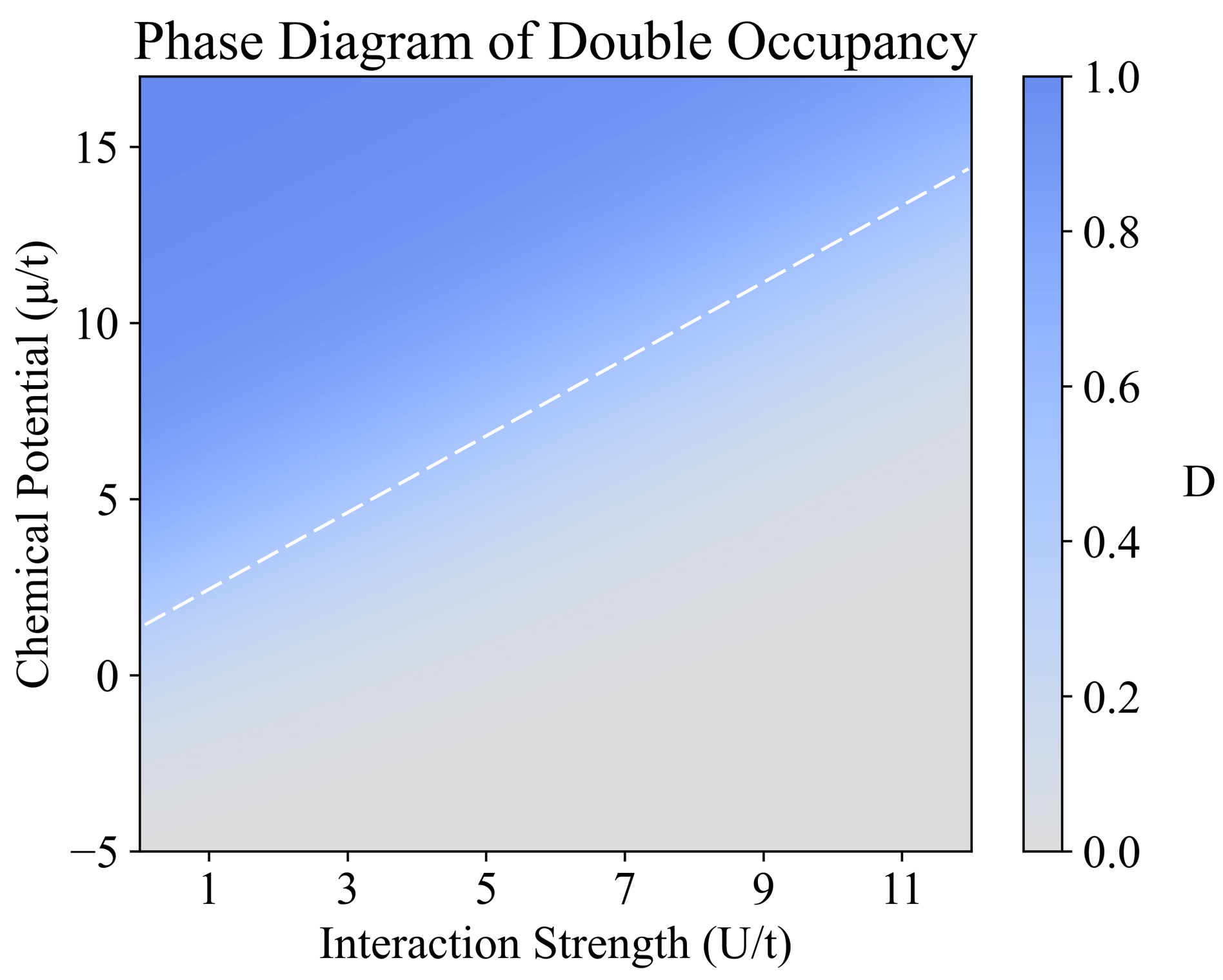}}%
    \subfigure[]{ \includegraphics[width=0.34\linewidth]{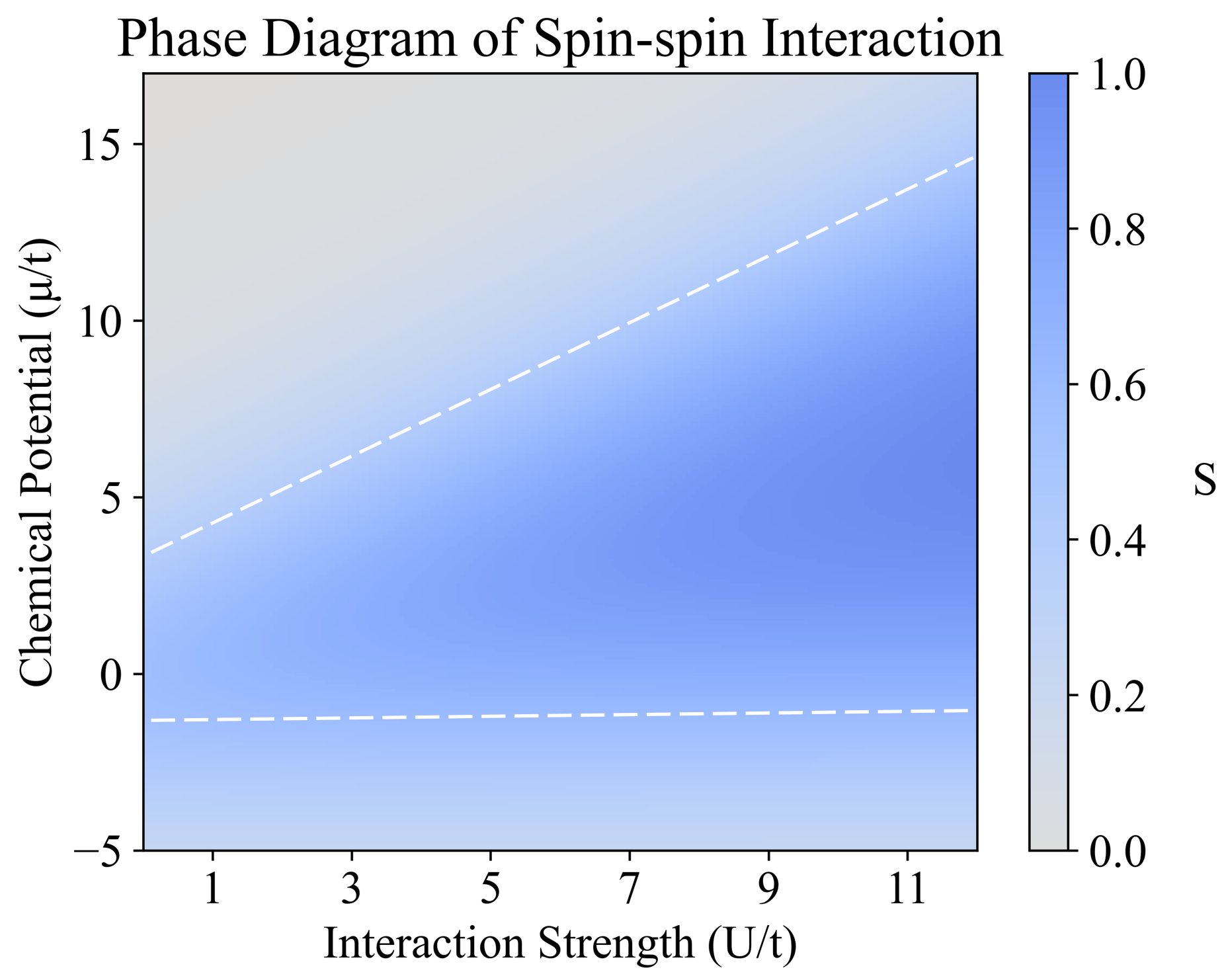}}
    \subfigure[]{ \includegraphics[width=0.68\linewidth]{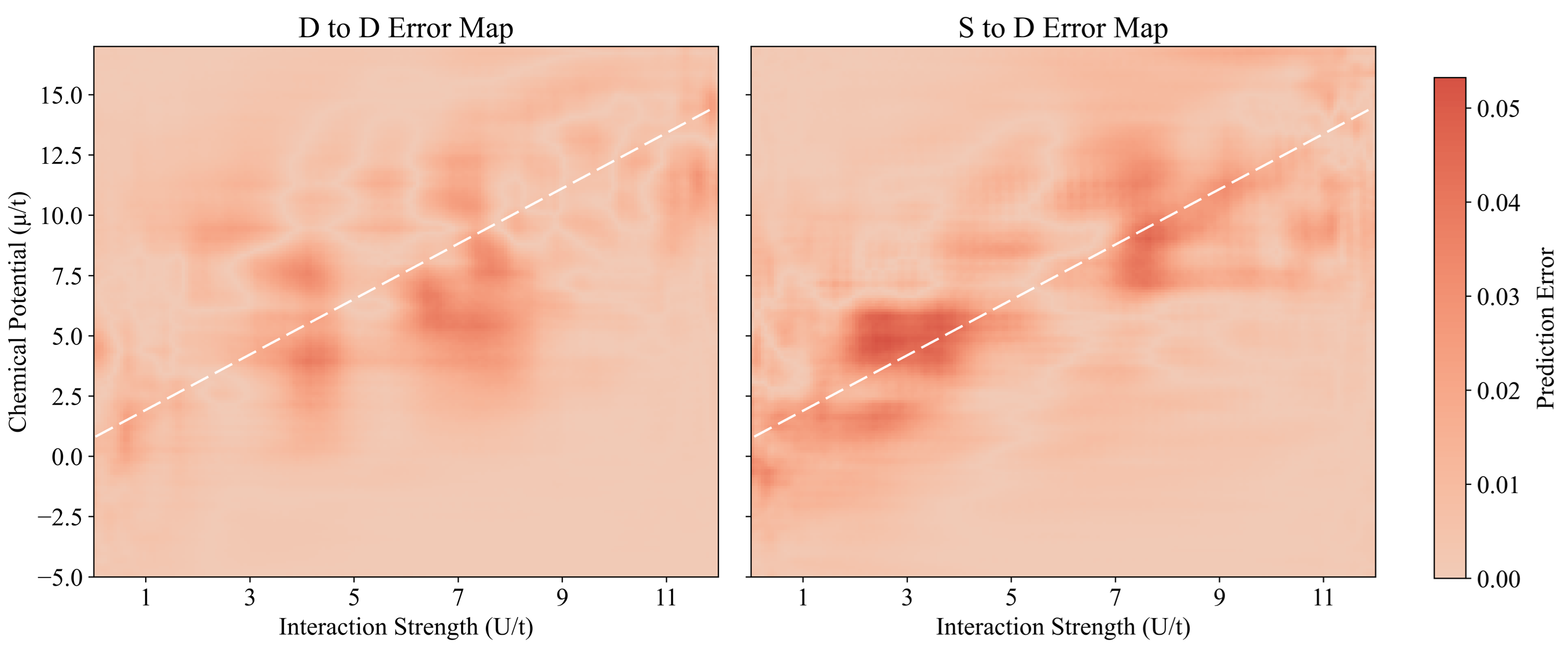}}
    \subfigure[]{ \includegraphics[width=0.68\linewidth]{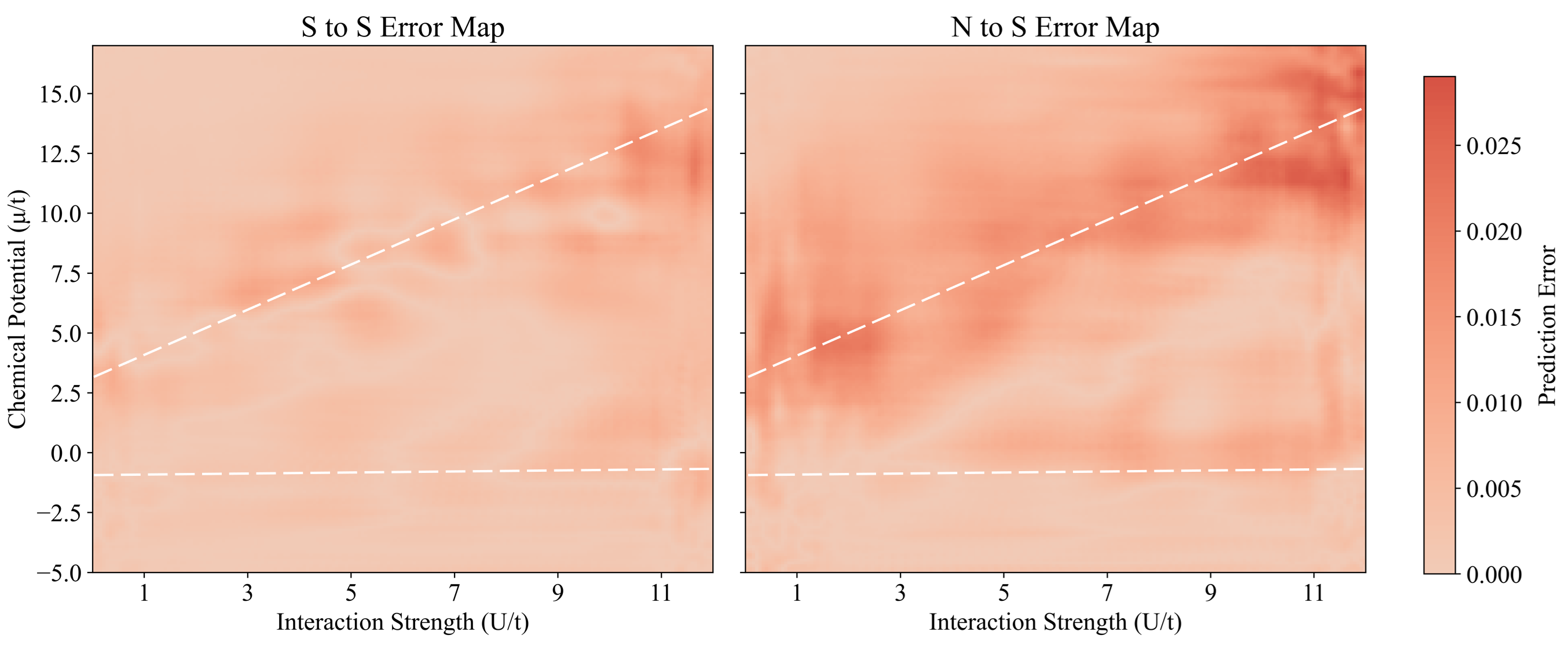}}
    \caption{Phase analysis of mapping error. Demonstration crossover lines are marked on phase diagrams of (a) double occupancy, and (b) spin-spin interaction. Same lines are also marked respectively on mapping error phase diagrams of (c) double occupancy, and (d) spin-spin interaction. A comparison of error phase distribution can be made in both (c) and (d) between self mapping and transferred mapping.}
    \label{phase_error}
\end{figure*}

In Figure \ref{phase_error} (a), we plot the double occupancy phase diagram of this piece of sample data. A sharp crossover region can be seen, indicating half-filling and a rapid emergence of non-trivial $D$ values. We put a white dashed line at the spot to indicate its location. In Figure \ref{phase_error} (c), we display phase-wise mapping errors from two mapping tests on this $D$ phase diagram in (a): self mapping "D to D" and transferred mapping "S to D". Error values here are shown in their absolute form, details can be seen in Methods. The redundancy considered lies in a comparison between these two error diagrams. Regions where transferred mapping does not show substantially larger error indicate high transferability and therefore strong redundancy, whereas regions where transferred mapping performs noticeably worse indicate reduced transferability. Figure \ref{phase_error} (c) shows that increased transferred error is concentrated near the crossover line, suggesting that information complexity increases in critical regimes, weakening redundancy and decoupling observables.

We perform a similar error analysis for the other information-flow direction using the spin-spin interaction phase diagram in Figure \ref{phase_error} (b). We also put white dashed lines to indicate locations of critical regimes, including the Mott crossover regime around half-filling. The corresponding error maps are displayed in Figure \ref{phase_error} (d). We again see sudden transferred error increase around critical regimes, indicating a sharp reduction of transferability and redundancy. These two tests together suggest that the exploitable redundancy considered here is strongest in regions with relatively smooth physical information. Within trivial phases, redundancy ensures a deep coupling between separate observables. Around critical regimes such as the Mott crossover or AFM onset regime however, meaningful physical information rises rapidly and diminishes this effect. 

\subsection{Extension across plaquettes}
A final worry comes from finite-size limitation. Previous sections have demonstrated that separate observables contain redundant physical information within a $2\times2$ Fermi-Hubbard plaquette. A central question remaining, however, is whether this redundancy is merely a property of the minimal plaquette geometry or whether it persists across plaquette boundaries when multiple plaquettes are connected to form a larger lattice. A comprehensive investigation of this issue is seriously hindered by the explosive computational cost to expand a fermion system. Empirically, lattice site number beyond $6$ or $7$ will be impractical to solve exactly. To address this obstacle in this sample test, we adopt the Finite-Temperature Lanczos Method~(FTLM) pipeline~\cite{FTLM} instead. Although generation of a large database remains challenging, FTLM can help to approximate sample data points of some larger Fermi-Hubbard systems without suffering from memory overflow. Meanwhile luckily, we do not necessarily need a new large training database because there is evidence showing that the model we have learned on the $2\times2$ lattice can be generalized to larger systems with satisfying accuracy given that governing physical laws remain unchanged~\cite{NNIQS}. Therefore, in this sample test, we will generate sample phase diagrams of some larger Fermi-Hubbard systems with FTLM operation, and evaluate transferability within these larger systems with existing models accordingly as a stringent test condition.

\begin{figure}
    \centering
    \includegraphics[width=1.0\linewidth]{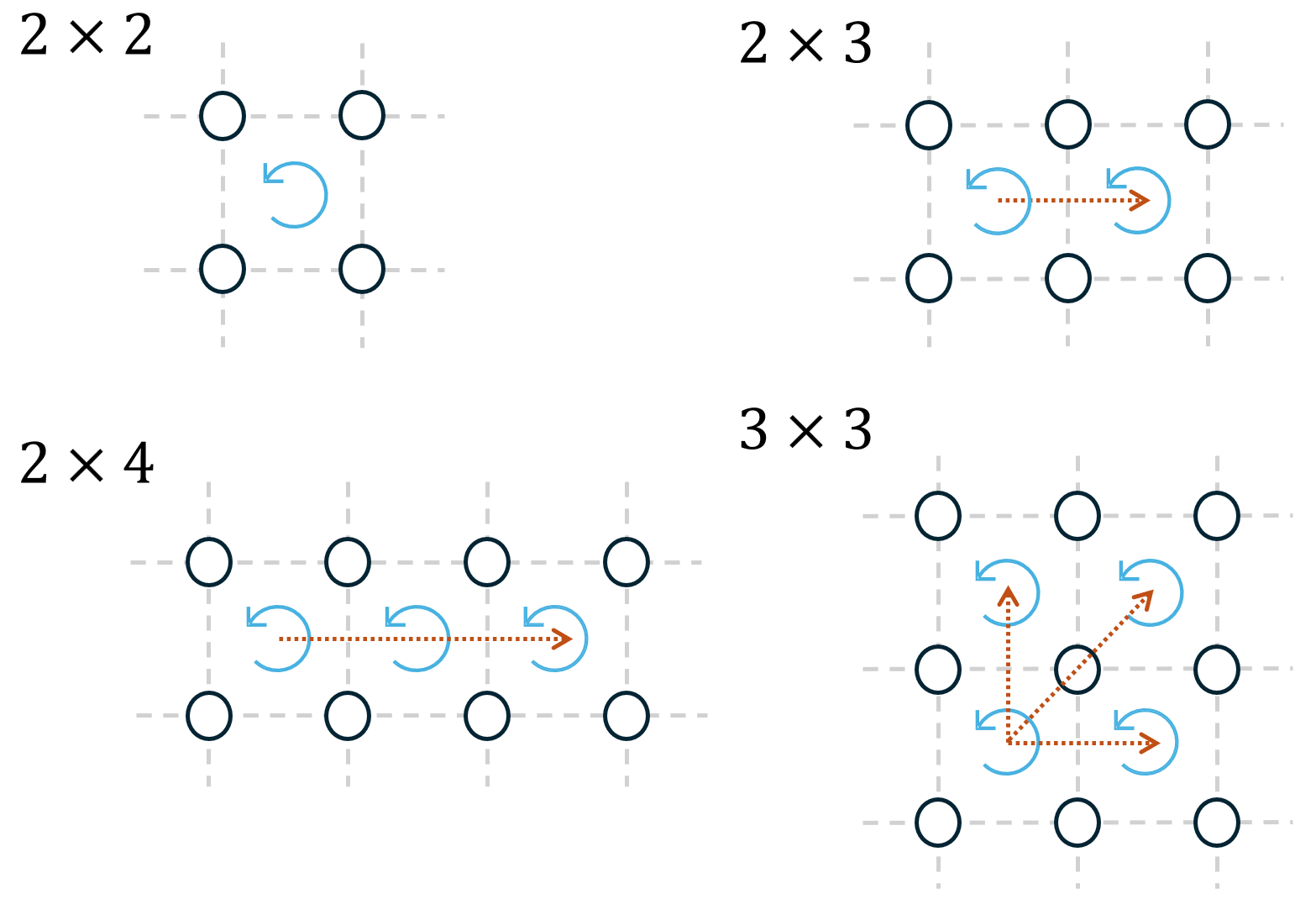}
    \caption{Longest lattice distances in sense of plaquettes. Except for $2\times2$ lattice, all three other lattices have their own forms of extension across plaquette boundaries.}
    \label{plaquettes}
\end{figure}

We will cover three different larger system architectures here, including $2\times3$, $2\times4$ ladders and a $3\times3$ square lattice respectively. The reason why we pick these three architectures lies within their plaquette structures. As we mentioned, the goal of this sample test is to verify whether redundancy can extend across plaquette boundaries, so we need our sample architectures to possess as many different boundary types as possible. A graph demonstration can be seen in Figure \ref{plaquettes}. In a $2\times2$ lattice, there is no inter-plaquette structures, while in all other lattices, there are some kinds of boundary-crossings. In a $2\times3$ lattice, we can go as far as to a neighboring plaquette, while in a $2\times4$ lattice, we can go further to the second nearest neighbor. In a $3\times3$ lattice, we can go simultaneously to two neighboring plaquettes in a 2D sense, as well as a diagonal plaquette which does not exist in other architectures. If redundancy proves to persist in all these architectures, they would together verify that the effects we have witnessed are not strictly local, they can generalize across plaquette boundaries beyond single-plaquette step and 1D direction.

\begin{figure*}
    \centering
    \includegraphics[width=1.0\linewidth]{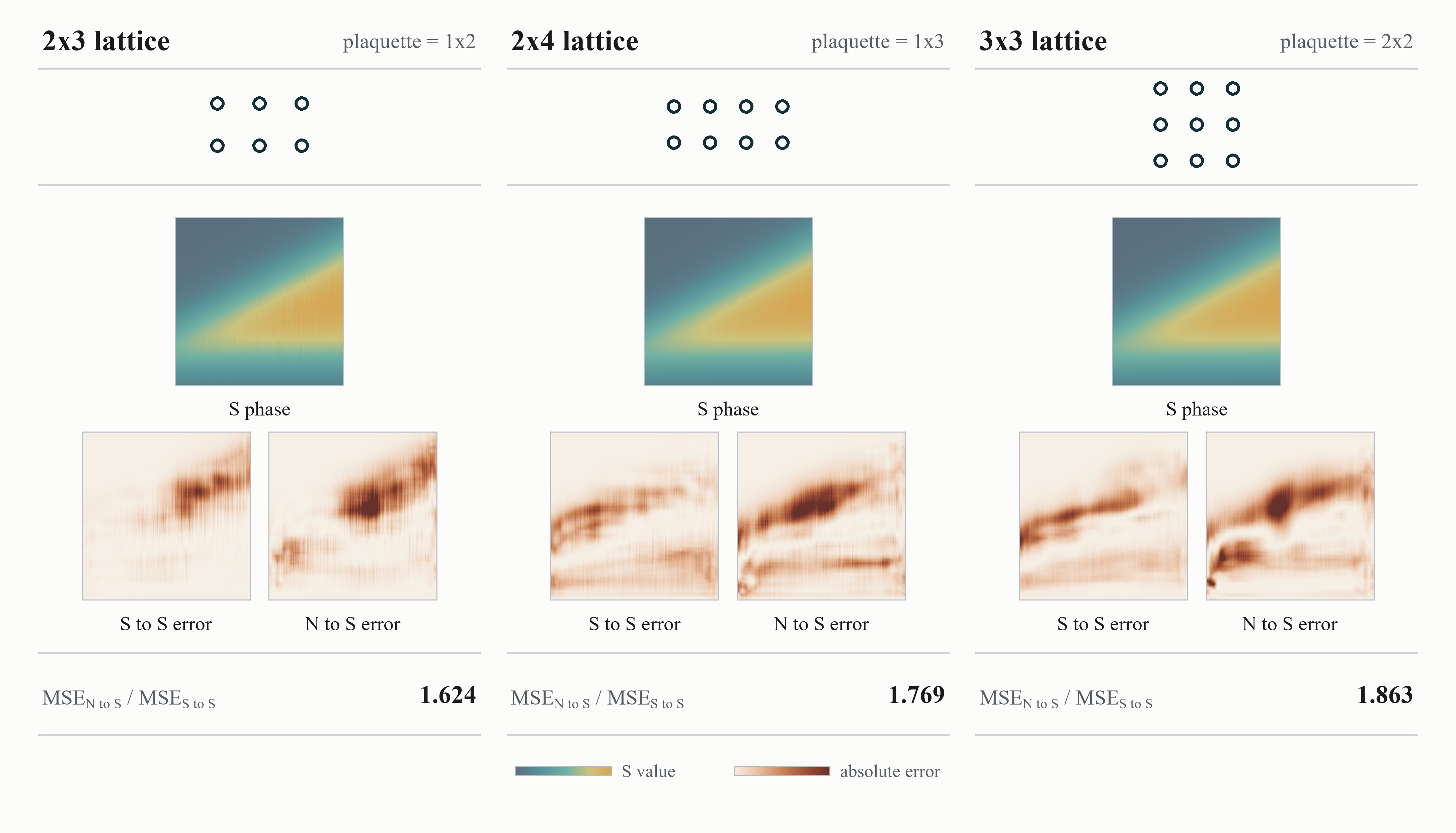}
    \caption{Large-system transferability test results. Different color scales are used to separate these results from previous ones. Minor vertical textures on the diagrams arise from unavoidable stochastic trace estimation during FTLM process and do not affect qualitative structures. "N to S" versus "S to S" mapping pair is chosen as the most representative comparison related to correlation length, and therefore lattice architecture. The corresponding statistical transferability in the form of MSE ratio has been shown for each of the architectures. Note that this value is a collective result from several sample diagrams and may not correspond precisely to the shown one.}
    \label{main_figure}
\end{figure*}

Test results have been shown in Figure \ref{main_figure}. We have chosen the "N to S" versus "S to S" mapping pair from Figure \ref{phase_error} as our test pair here, because while total density is fully local, staggered spin-spin correlation greatly involves global interactions. Therefore, this comparison pair best represents possible impacts on transferability from enlarging lattice architectures and thus increasing correlation lengths. We follow the FTLM workflow to sample large systems within a well-behaved thermodynamic range and collect statistical transferability data following same logic as in previous sections. Details can be seen in Methods. To understand these results, firstly, it is noticeable how similar all three S phase diagrams here are to the one we have shown in Figure \ref{phase_diagrams} (c), which is another justification to use our trained models in this test. Secondly, it can be seen from the three MSE ratios that we do not witness a drastic increase of transferred error compared to self-mapping when system size and plaquette number grow. As a benchmark, the same ratio reads $1.72$ in $2\times2$ lattice in Table \ref{psnr_matrix} (b). This indicates that transferability, and therefore physical redundancy is indeed generalizable across plaquette boundaries beyond single-plaquette step and 1D direction. Lastly, across all three architectures, the error phase diagrams also show stable and similar trends as we seen before, with transferred error increase mainly concentrated around phase critical regimes. This indicates that the break down mechanism of physical redundancy is also non-local.

\section{Discussion}
From the theory side, it is physically plausible that quantum many-body systems may contain redundancy among different observables, enabling transferred inference between them. Our results however suggest that the extent of this effect can be substantial: when transferred reconstructions approach comparable accuracy level of self-reconstructions in both information-flow directions, the observables may need to be interpreted as different lower-order projections of a shared higher-order many-body manifold. Further studies are needed to clarify the mechanism and limiting factors of this effect.

From the experimental side, after these results, we may need to reconsider our understanding on modern quantum observations. In many quantum systems, the measurement cost of separate physical observables can differ remarkably. As good examples, total density evaluation can be easily done in cold atom systems or doped condensed matter systems by techniques like STM. However, some other observables like double occupancy can be much harder to evaluate, and in many cases require indirect inference~\cite{discussion_1, discussion_2, discussion_3}. If physical redundancy can be identified and exploited in these systems, it is very likely that we can bypass the most challenging parts of observation by transferred inference in experiments that do not necessarily require ultra-high precision.

We should also stress here that although we find evidence of strong redundancy generalizability across plaquette boundaries, our study stays at finite-size due to complexity of the system. A future numerical study approaching thermodynamic limit will be greatly helpful to refine this effect in a bulky context.

\section{Conclusion and Outlook}
In this work, we investigate physical redundancy in the Fermi-Hubbard system by conducting transferability tests among three representative observables: total density, double occupancy, and spin-spin correlation. Although these observables probe different physical aspects of the system and belong to different information hierarchies, we find unexpectedly strong transferability between them, indicating substantial observable-level redundancy. Further tests show that this redundancy persists across the investigated temperature range and weakens mainly in critical regimes, where phase structures become sharper and intrinsic physical information changes rapidly. Same behaviors are witnessed in both within-plaquette and across-plaquette scenarios.

Future theoretical and numerical studies will be important for revealing the mechanism of this effect and quantifying its limiting factors. Such understanding may help exploit physical redundancy in quantum simulation and provide further insight into the many-body information structure behind correlated quantum systems.

\section{Methods}
\begin{figure}
    \centering
    \includegraphics[width=1.0\linewidth]{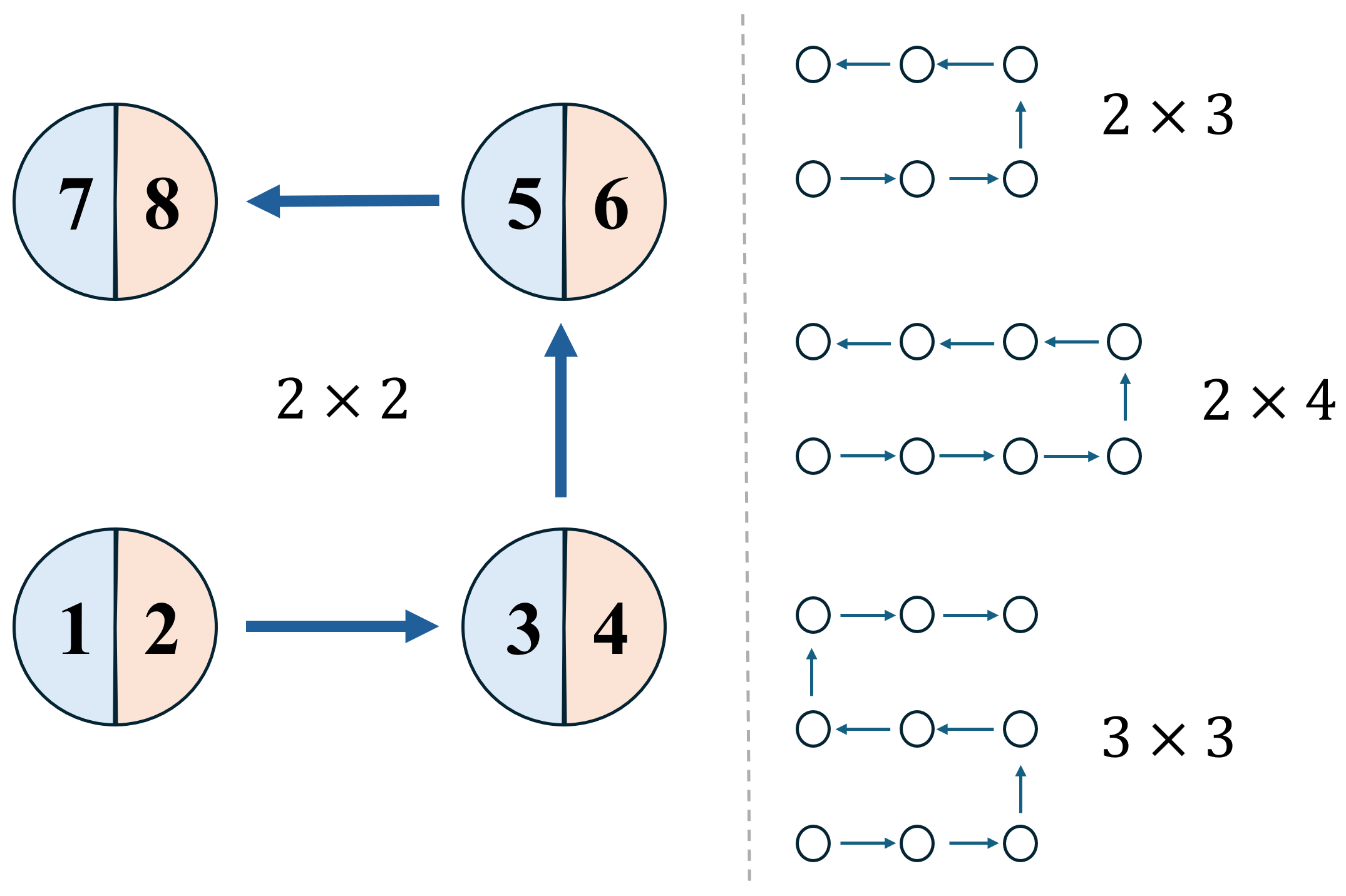}
    \caption{Orderings of modes. $2\times2$ architecture is shown in details, while the others follow a similar design. A blue hemisphere is used to represent an up-spin mode, while an orange hemisphere is used to represent a down-spin mode. An "S" shape ordering is adapted to minimize the need for series multiplication.}
    \label{ordering}
\end{figure}

There is some freedom in ordering fermionic modes due to basis-change symmetry~\cite{order} in Jordan-Wigner transformation. The orderings used in this work are shown in Figure \ref{ordering}. Our choice has two advantages. Firstly, it keeps interacting modes as close as possible, although an $(i,j)$ pair may still correspond to second-nearest interaction. In physical implementations, this advantage can greatly reduce setup difficulty. Secondly, the system Hamiltonian contains strings of Pauli operators after Jordan-Wigner transformation, and our orderings here help minimize the length and frequency of such strings.

During generation of our central dataset, parameter $t$ in the system Hamiltonian is fixed to be $1$, while other parameters are set with respect to it. In order to thermodynamically cover different system characteristics~\cite{theory1}, we scan temperature $T$ from $T/t = 0.01$ to $T/t = 10$ to cover all scenarios from quantum dominance to thermal dominance. At each specific temperature $T$, a phase diagram for each of the three observables are generated. In each phase diagram, $U/t$ and $\mu/t$ are used as X and Y axis correspondingly. $U/t$ ranges from $[0,12]$ with 192 equally separated values, while $\mu$ ranges from $[-5,17]$ with the same number of distinct values. These ranges are designed to cover as more different physical regions on phase diagrams as possible, including various density of states within both metal and Mott insulator phase~\cite{theory1, theory3}. With more comprehensive physics included, more solidity of the model can be verified. For our sample test with larger architectures however, due to increasing complexity, we only sample within $T/t \in [1.5,3]$, which is a well-behaved range with clear phase structures and well-balanced contributions from different terms within system Hamiltonian. The statistical numbers we present in Figure \ref{main_figure} are collective results respectively from all sample diagrams we generate within this range.

Each data point in the dataset is taken as the sigmoid value by the sigmoid function
\begin{equation}
    SigmoidValue = \frac{1}{1+e^{-RealValue}}
\end{equation}
This measure helps to re-scale the estimated expected values between [0,1] to increase convergence during training. However, in some of our tests, in order to display results in a more engineering-friendly way, we have converted the values back to real ones by a reverse operation of sigmoid function. MSE values seen in the T-scan experiment and absolute values seen in the error phase analysis are both displayed in this way. $n$ values in the noise test are also managed in this way, because self-proportional Gaussian noises and $(0,2)$ clipping is only meaningful in real values.

The numerical object we use to accommodate a piece of data during training is a 3D array. On dimension $0$, there are three layers, each holding a phase diagram from one observable. Therefore the original architecture of an object include all $(A,B,C)$ observable information. When we try to do a "from A map to B" transferability test, we want our input object to keep only A information, and our ground truth object to keep both A and B information~(as discussed above, A information in ground truth is used for supervision). In other words, we need to wipe out B, C information in the input object and C information in the ground truth object. To do that, we manually set these layers to a plain surface of value $0.5$. This value corresponds to a $0$ value prior to sigmoid function, and is optimal for overall convergence. It can be easily seen from mapping results which we do not put in the Results section that this arrangement do not interfere with our task. After cleaning our input object and keeping only A information, we still need a clipping operation because in order to motivate the network to do mapping, we cannot allow full information of A input into it. Ideally, only part of A's information should be provided, and the network will try to map to B alongside reconstructing A's self information. In a random clipping and batched training setup, the network will gradually see all of A's information, but the clipping will make sure that it does not see everything altogether at once.

\bibliography{reference}
\end{document}